\documentclass[11pt]{article}
\usepackage{graphicx}
\title{\bf Solutions to the Modified P\"{o}schl-Teller potential in $D$-dimensions.}
\author{\bf D. Agboola\footnote{tomdavids2k6@yahoo.com}}
\date{Department of Pure and Applied Mathematics,\linebreak Ladoke Akintola University of Technology,Oyo State, Nigeria. \linebreak  P.M.B. 4000.}\em

\begin{document}
\maketitle
\vspace{0.5in}
\noindent {\bf Abstract:} An approximate solution of the $D$-dimensional Schr$\ddot{o}$dinger equation with the modified P$\ddot{o}$schl-Teller potential is obtained with an approximation of the centrifugal term. Solution to the corresponding hyper-radial equation is given using the conventional Nikiforov-Uvarov method. The normalization constants for the P$\ddot{o}$schl-Teller potential are also computed. The expectation values $\langle r^{-2}\rangle$,$\langle V(r)\rangle$, are also obtained using the Feynman-Hellmann theorem.\\

\vspace{0.5in}
\vspace{1in}


\pagebreak
\noindent { \bf 1.0 Introduction}\\

The exact solutions of the Schr$\ddot{o}$dinger equation is only obtainable for a few number of physical potentials. However, the P$\ddot{o}$schl-Teller potential happens to be one of the few potentials that enjoy an exact solution of the Schr$\ddot{o}$dinger equation. Over the past years, a number of researchers have worked on the P$\ddot{o}$schl-Teller potential [1-5, 11-16, 24]. For instance, a unified treatment of the quasi-exactly solvable P$\ddot{o}$schl-Teller potential was discussed by Koc and Koca [15]. The Casimir operator of $SO(1,2)$ and the P$\ddot{o}$schl-Teller potential was presented by Capelas de Oliveira and da Rocha [13] using the anti-de sitter(Ads) space time approach. The complex and $PT$- symmetric non-Hermitian P$\ddot{o}$schl-Teller potential has also been discussed by  Yesiltas and coworkers [9]. Moreover, the modified P$\ddot{o}$schl-Teller potential as been used to derive the well-known $SO(2)$ spectrum generating algebra for an infinite square well problem [44]. Recently, the supersymmetric solution of the modified P$\ddot{o}$schl-Teller was discussed by Aktas and Sever [2], and the supersymmetric modified P$\ddot{o}$schl-Teller and delta-well potential has been studied by Diaz {\it et al} [5].

The extension of physical problems to higher dimensional spaces plays an important role in many area of physics. Some of the physical system of interest in quantum mechanics which have been thoroughly studied in $N$-dimensional spaces are the two exactly solvable potentials, which are, the  Coulomb and harmonic oscillator potentials [28-41]. In particular, Oyewumi and Ogunsola [19] have consider the exact solutions of the harmonic oscillator in multidimensional spaces. Recently, Oyewumi [18] presented the analytic solutions of the Kratzer-Fues potential in arbitrary dimensions, and also the exactly complete solutions of the pseudoharmoinic potential in $N$-dimensions has been discussed by Agboola {\it et al} [17].

It is therefore the aim of this work to present a multi-dimensional treatment of the modified P$\ddot{o}$schl-Teller within the framework of an approximation to the centrifugal term. The paper is arranged as follows: the next section gives the Schr$\ddot{o}$dinger equation in
$D$-dimensions; section 3 gives a brief description of the Nikiforov-Uvarov
method, while the following section presents the bound-state solutions to the P$\ddot{o}$schl-Teller potential. In section 5, some expectation values are calculated; and finally,the conclusions of the work are presented in section 6.

\pagebreak
\noindent{\bf 2.~The Schr$\ddot{o}$dinger Equation with the Modified P$\ddot{o}$schl-Teller \linebreak Potential in $D$-dimensions.} \\
We consider the Schr$\ddot{o}$dinger equation [17-21]
$$H\Psi_{n\ell m}(r,\Omega)=E\Psi_{n\ell m}(r,\Omega),  \eqno{(1)}$$
where $E$ is the energy eigenvalue,$\Psi_{n\ell m}(r,\Omega)$ is the wave function and $H$ is the Hamiltonian given by
$$H=-\frac{\hbar^2}{2\mu}\left[r^{1-D}\frac{\partial}{\partial r}\left(r^{D-1}\frac{\partial}{\partial r}\right)+\frac{\Lambda_D^2}{r^2}\right]+V(r),  \eqno{(2)}$$
$\mu$ is the reduced mass, $\hbar$ the Planck's constant and $V(r)$ is the modified P$\ddot{o}$schl-Teller potential given as [2,5,11]
$$V(r)=-\frac{V_0}{\cosh^2(\alpha r)},  \eqno{(3)}$$
$\alpha$ being a parameter and $V_0$ is the potential depth.\\
Inserting (2)and (3) into (1) and separating the variables as follows
$$\Psi_{n\ell m}(r,\Omega)=r^{-(D-1)/2}R_{n_r,\ell}(r)Y_\ell^m(\Omega),  \eqno{(4)}$$ 
Eq.(1) reduces to two separate equations
$$R_{n_r\ell}^{\prime\prime}(r)+\left[\frac{2\mu}{\hbar^2}\left(E+\frac{V_0}{\cosh^2(\alpha r)}\right)-\frac{(k-1)(k-3)}{4r^2}\right]R_{n_r\ell}(r)=0,  \eqno{(5)}$$
with $k=D+2\ell$, and 
$$\Lambda_D^2Y_\ell^m(\Omega)+\ell(\ell+D-2)Y_\ell^m(\Omega)=0, \eqno{(6)}$$
where $Y_\ell^m(\Omega)$ are the hyperspherical harmonics with
$$\small \Lambda_D^2=\sum_{i=1}^{D-2}\left(\prod_{j=0}^i\sin\theta_j\right)^{-2}(\sin\theta_i)^{i+3-D}\frac{\partial}{\partial \theta_i}\left(\sin\theta_i^{D-i-1}\frac{\partial}{\partial\theta_i}\right)+\left(\prod_{j=1}^{D-2}\sin\theta_j\right)^{-2}\frac{\partial^2}{\partial^2\phi} \eqno{(7)}$$
and the separation constant
$$\beta=\ell(\ell+D-2), \hspace{1cm}\ell=0,1,2,... \eqno{(8)}$$
Equation (5) is the hyperradial  equation for the modified P$\ddot{o}$schl-Teller \linebreak potential in $D$-dimensions, $n_r$ the hyperradial quantum number and $\ell$ is the orbital angular momentum quantum number.\\\\
\noindent {\bf 3. The Nikiforov-Uvarov method.}\\\\
In this section, we give a brief description of the conventional Nikiforov-Uvarov method. A more detailed description of the method can be obtained the following references [20].With an appropriate transformation $s=s(r)$,the one dimensional Schr$\ddot{o}$dinger equation can be reduced to a generalized equation of hypergeometric type which can be written as follows:
$$\psi^{\prime\prime}(s)+ \frac{\tilde{\tau}(s)}{\sigma(s)}\psi^\prime(s)+ \frac{\tilde{\sigma}(s)}{\sigma^2(s)}\psi(s)=0  \eqno{(9)}$$ 
Where $\sigma(s)$and $\tilde{\sigma}(s)$ are polynomials, at most second-degree, and $\tilde{\tau}(s)$is at most a first-order polynomial. To find particular solution of Eq. (9) by separation of variables, if one deals with
$$\psi(s)=\phi(s)y_{n_r}(s),  \eqno{(10)}$$
Eq.(9)becomes
$$\sigma(s)y^{\prime\prime}_{n_r}+\tau(s)y^\prime_{n_r} +\lambda y_{n_r} =0  \eqno{(11)}$$
where
$$\sigma(s)= \pi(s)\frac{\phi(s)}{\phi^\prime(s)},  \eqno{(12)}$$,
$$\tau(s)=\tilde{\tau}(s)+2\pi(s) ,  \tau^\prime(s)<0,  \eqno{(13)}$$\\
$$\pi(s)=\frac{\sigma^\prime-\tilde{\tau}}{2}\pm \sqrt{\left(\frac{\sigma^\prime-\tilde{\tau}}{2}\right)^2-\tilde{\sigma}+t\sigma},  \eqno{(14)}$$
and 
$$\lambda=t+\pi^\prime(s).  \eqno{(15)}$$
The polynomial $\tau(s)$ with the parameter $s$ and prime factors show the differentials at first degree be negative. However,determination of parameter $t$ is the essential point in the calculation of $\pi(s)$. It is simply defined by setting the discriminate of the square root to zero [20]. Therefore, one gets a general quadratic equation for $t$.The values of $t$ can be used for calculation of energy eigenvalues using the following equation
$$\lambda=t+\pi^\prime(s)=-n_r\tau^\prime(s)-\frac{n_r(n_r-1)}{2}\sigma^{\prime\prime}(s).   \eqno{(16)}$$
Furthermore, the other part $y_{n_r}(s)$ of the wave function in Eq.(12) is the hypergeometric-type function whose polynomial solutions are given by Rodrigues relation: 
$$y_{n_r}(s)=\frac{B_{n_r}}{\rho(s)}\frac{d^{n_r}}{ds^{n_r}}[\sigma^{n_r}(s)\rho(s)]  \eqno{(17)}$$ 
where $B_{n_r}$ is a normalizing constant and the weight function $\rho(s)$ must satisfy the condition [20]
$$(\sigma\rho)^\prime =\tau\rho.   \eqno{(18)}$$\\\\
\noindent{\bf 4.0 Eigenvalues for the modified P$\ddot{o}$schl-Teller potential.}\\
For small $\alpha$, one can use the approximation of the centrifugal potential given as $$\frac{1}{r^2}\approx\frac{\alpha^2}{\sinh^2(\alpha r)}.\eqno(19)$$ Thus Eq.(5) becomes $$R^{\prime\prime}_{{n_r}\ell}(r)+\left[\frac{2\mu}{\hbar^2}\left(E+\frac{V_0}{\cosh^2(\alpha r)}\right)-\frac{(k-1)(k-3)\alpha^2}{\sinh^2(\alpha r)}\right]R_{{n_r}\ell}(r)=0   \eqno{(20)}$$
If we take a transformation $s=\tanh^2(\alpha r)$, after some straight forward manipulations, Eq.20 becomes 
$$R_{n_r\ell}^{\prime\prime}(s)+\frac{1-3s}{2s(1-s)}R_{n_r\ell}^\prime(s)+\frac{1}{4s^2(1-s)^2}[-\delta s^2+(\gamma+\delta-\epsilon^2)s-\gamma]R_{n_r\ell}(s)=0  \eqno{(21)}$$
where$$ -\epsilon^2=\frac{2\mu E}{\hbar^2\alpha^2},\hspace{.1in} \delta=\frac{2\mu V_0}{\hbar^2\alpha^2}\hspace{.1in}  \mbox{and}\hspace{.1in}  \gamma=\frac{1}{4}(k-1)(k-3).  \eqno{(22)}$$ 
Comparing Eqs. (21) and (9) we can define the following:\\
$$\tilde{\tau}(s)=1-3s\hspace{.1in},\sigma(s)=2s(1-s)\hspace{.1in} \mbox{and}\hspace{.1in}\tilde{\sigma}(s)=-\delta s^2+(\gamma+\delta-\epsilon^2)s-\gamma \eqno{(23)}$$
Inserting these into Eq.(14), we have the following function $$\pi(s)=\frac{1-s}{2}\pm\frac{1}{2}\sqrt{(1+4\delta-8t)s^2+(8t-4(\gamma+\delta-\epsilon^2)-2)s+4\gamma+1}  \eqno{(24)}$$ The constant parameter $t$ can be found by the condition that the discriminant of the expression under the square root  has a double root, i.e., its discriminant is zero. Thus the possible value function for each value of $t$ is given as 
$$\pi(s)=\frac{1-s}{2}\pm \left\{\begin{array}{lll}
\frac {1}{2}\left [\left(-2\epsilon +\sqrt{1+4\gamma}\right)s-\sqrt{1+4\gamma}\right] & \mbox{for} & t=-\frac{1}{2}(\gamma-\delta+\epsilon^2)+\frac{1}{2}\epsilon\sqrt{1+4\gamma}\\\\
\frac {1}{2}\left[\left(2\epsilon +\sqrt{1+4\gamma}\right)s-\sqrt{1+4\gamma}\right] & \mbox{for} & t=-\frac{1}{2}(\gamma-\delta+\epsilon^2)-\frac{1}{2}\epsilon\sqrt{1+4\gamma}\end{array}\right. \eqno{(26)}$$\\
By Nikiforov-Uvarov method, we made an appropriate choice of the function $\pi(s)=\frac{1-s}{2}-\frac {1}{2}\left[\left(2\epsilon +\sqrt{1+4\gamma}\right)s-\sqrt{1+4\gamma}\right]$ such that by Eq.(16), we can obtain the eigenvalue equation to be $$-\frac{1}{2}(\gamma-\delta+\epsilon^2)-\frac{1}{2}\epsilon\sqrt{1+4\gamma}-\frac{1}{2}(2\epsilon+\sqrt{1+4\gamma})-\frac{1}{2}=n_r[4+2\epsilon+\sqrt{1+4\gamma}]+2n_r(n_r-1)  \eqno{(27)}$$ Eq.(27) can be written in the powers of $\epsilon$ as follows
$$\epsilon^2+\epsilon\left[2(2n_r+1)+\sqrt{1+4\gamma}\right]+(\gamma-\delta)+\left[(1+2n_r)+\sqrt{1+4\gamma}\right]=0,  \eqno{(28)}$$	
such that we can obtain the solution as 
$$-\epsilon^2=-\frac{1}{4}\left[-2(2n_r+1)-\sqrt{1+4\gamma}+\sqrt{1+4\delta}\right]^2.  \eqno{(29)}$$
Using the relation in Eq.(22) and $k=D+2\ell$, we have the energy eigenvalue as
$$E_{n_r\ell}=-\frac{\hbar^2\alpha^2}{8\mu}\left[-2\left(2n_r+\ell+\frac{D}{2}\right)+\sqrt{1+\frac{8\mu V_0}{\hbar^2\alpha^2}}\right]^2  \eqno{(30)}$$
One can define the principal quantum number $n=2n_r+\ell$, such that Eq.(30) becomes 
$$E_{n}=-\frac{\hbar^2\alpha^2}{8\mu}\left[-\left(2n+D\right)+\sqrt{1+\frac{8\mu V_0}{\hbar^2\alpha^2}}\right]^2,n=0,1,2,...  \eqno{(31)}$$
We now give the following comments: Eq.(31) is the approximate energy spectrum of the P$\ddot{o}$schl-Teller potential in $D$-dimensions.However, in order to check the validity of the approximation (19), if we consider the $s-$wave and let $D=1$, we have the equation 
$$E_{n}=-\frac{\hbar^2\alpha^2}{8\mu}\left[-\left(2n+1\right)+\sqrt{1+\frac{8\mu V_0}{\hbar^2\alpha^2}}\right]^2\hspace{0.1in}n=0,1,2,...  \eqno{(32)}$$ which is consistent with the results obtained in previous works [2,24]. Also from Eq.(31), one can obtain the critical screening parameter, $\alpha_c$ (for which $E_n=0$) as $$\alpha_c=\sqrt{\frac{8\mu V_0}{\hbar^2[(2n+D)^2-1]}}.  \eqno{(33)}$$ Finally, the curves of the energy levels $E_n$ against $\alpha$ for some dimensions (figure 1) reveals that the values of the energy for some adjacent dimensions becomes closer for higher exited states.   \\ 

\noindent{\bf 4.1 Eigenfunctions for the modified P$\ddot{o}$schl-Teller potential.}\\\\
In this section, we obtain the wave functions using the Nikiforov-Uvarov method. By substituting $\pi(s)$ and $\sigma(s)$ into Eq.(12), and solving the first order differential  equation to have
$$\phi(s)=s^{2v}(1-s)^{\epsilon/2}, \hspace{0.5in} 2v=2\ell+D-1.  \eqno{(34)}$$
Also by Eq.(18), the weight function $\rho(s)$ can be obtained as 
$$\rho (s)=\frac{1}{2}s^{v}(1-s)^\epsilon  \eqno{(35)}$$
Substituting Eq.(35)into the Rodrigues relation (17), we have
$$y_{n_r}(s)=B_{n_r}s^{-v}(1-s)^{-}\frac{d^{n_r}}{ds^{n_r}}\left[s^{n_r+v}(1-s)^{n_r+\epsilon}\right]. \eqno{(36)}$$ 
Therefore, we can write the wave function $R_{{n_r}\ell}(s)$ as 
$$R_{{n_r}\ell}(s)=C_{n_r}s^{2v}(1-s)^{\epsilon/2}P^{(v,\epsilon)}_{n_r}(1-2s)  \eqno{(37)}$$
where $C_{n_r}$ is the normalization constant, and we have used the definition of the Jacobi polynomials[26],given as
$$P^{(a,b)}_n(s)=\frac{(-1)^n}{n!2^n(1-s)^a(1+s)^b}\frac{d^n}{ds^n}\left[(1-s)^{a+n}(1+s)^{b+n}\right].  \eqno{(38)}$$
The curves of the unnormalized hyperradial function are plotted against $r$ for some $\ell\neq 0$  states in figures 4-7.

Moreover, to compute the normalization constant $C_{n_r}$, it is easy to show that $$\int^\infty_0\left|r^{\frac{-(D-1)}{2}}R_{{n_r}\ell}(r)\right|^2r^{D-1}dr=\int^\infty_0|R_{{n_r}\ell}(r)|^2dr=\int^1_0|R_{{n_r}\ell}(s)|^2\frac{ds}{2\alpha\sqrt{s}(1-s)}=1  \eqno{(39)}$$
where we have also used the substitution $s=\tanh^2(\alpha r)$. Putting Eq.(37) into Eq.(39) and using the following definition of the Jacobi polynomial[26]
$$P^{(a,b)}_n(s)=\frac{\Gamma(n+a+1)}{n!\Gamma(1+a)} \ _2F_1\left(-n,a+b+n+1;1+a;\frac{1-s}{2}\right),  \eqno{(40)}$$ we arrived at 
$$C_{n_r}^2 N_{n_r}\int_0^1s^{4v-\frac{1}{2}}(1-s)^{\epsilon-1}[\ _2F_1\left(-n_r,\epsilon+v+n_r+1;1+v;s\right)]^2ds=\alpha \eqno{(41)}$$ where $N_{n_r}=\frac{1}{2}\left[\frac{\Gamma(n_r+v+1)}{n_r!\Gamma(v+1)}\right]^2$ and $_2F_1$ is the hypergeometric function. Using the following series representation of the hypergeometric function
$$_pF_q(a_1,...,a_p;c_1,...,c_q;s)=\sum_{n=0}^\infty\frac{(a_1)_n...(a_p)_n}{(c_1)_n...(c_q)_n}\frac{s^n}{n!}  \eqno{(42)}$$we have
$$\small C_{n_r}^2 N_{n_r}\sum^{n_r}_{k=0}\sum^{n_r}_{j=0}\frac{(-n_r)_k(\epsilon+v+n_r+1)_k}{(v+1)_k k!}\frac{(-n_r)_j(\epsilon+v+n_r+1)_j}{(v+1)_j j!}\int_0^1s^{4v+k+j-\frac{1}{2}}(1-s)^{\epsilon-1} ds=\alpha .\eqno{(43)}$$
Hence, by the definition of the Beta function, Eq.(43)becomes
$$\small C_{n_r}^2 N_{n_r}\sum^{n_r}_{k=0}\sum^{n_r}_{j=0}\frac{(-n_r)_k(\epsilon+v+n_r+1)_k}{(v+1)_k k!}\frac{(-n_r)_j(\epsilon+v+n_r+1)_j}{(v+1)_j j!} B\left(4v+k+j+\frac{1}{2},\epsilon\right)=\alpha.  \eqno{(44)}$$
Using the relations $B(x,y)=\frac{\Gamma(x)\Gamma(y)}{\Gamma(x+y)}$ and the Pochhammer symbol \linebreak $(a)_n=\frac{\Gamma(a+n)}{\Gamma(a)}$, Eq.(44) can be written as \\
$$\small C_{n_r}^2 N_{n_r}\sum^{n_r}_{k=0}\frac{(-n_r)_k(\epsilon+v+n_r+1)_k(4v+\frac{1}{2})_k}{(\epsilon+4v+\frac{1}{2})_k (v+1)_k k!}\sum^{n_r}_{j=0}\frac{(-n_r)_j(\epsilon+v+n_r+1)_j(4v+k+\frac{1}{2})_j}{(\epsilon+4v+k+\frac{1}{2})_j (v+1)_j j!}=\frac{\alpha}{B(4v+\frac{1}{2},\epsilon)}  \eqno{(45)}$$

\noindent Lastly, Eq.(45) can be used to compute the normalization constants for $n_r=0,1,2,...$ In particular for the ground state, i.e $n_r=0$, we have
$$C_0=\sqrt{\frac{2\alpha}{B(4v+\frac{1}{2},\epsilon)}}  \eqno{(46)}$$

\pagebreak
\noindent {\bf 5. Some Expectation Values of the P$\ddot{o}$schl-Teller Potential in \linebreak $D$-dimensions.}\\\\
 We now calculate some expectation values of the P$\ddot{o}$schl-Teller potential using the Hellmann-Feynmann theorem (HFT) [27, 28]. Suppose the Hamiltonian $H$ for a particular quantum system is a function of some parameters $q$, and let $E(q)$ and $\Psi(q)$ be the eigenvalues and eigenfunctions of $H(q)$ respectively, then the HFT states that $$\frac{\partial E(q)}{\partial q}=\langle\Psi(q)|\frac{\partial H(q)}{\partial q}|\Psi(q)\rangle.  \eqno{(47)}$$ The effective Hamiltonian of the hyperradial function is given as $$H=\frac{-\hbar^2}{2\mu}\frac{d^2}{dr^2}+\frac{\hbar^2}{2\mu}\frac{(2\ell+D-1)(2\ell+D-3)}{4r^2}-\frac{V_0}{\cosh^2(\alpha r)}.  \eqno{(48)}$$
In order to calculate $\langle r^{-2}\rangle$, we set $q=\ell$ such that $$\langle\Psi(\ell)|\frac{\partial H(\ell)}{\partial \ell}|\Psi(\ell)\rangle=\frac{\hbar^2}{2\mu}(2\ell+D-2)\langle r^{-2}\rangle  \eqno{(49)}$$
and 
$$\frac{\partial E_{n_r\ell}(\ell)}{\partial \ell}=\frac{\hbar^2\alpha^2}{2\mu}\left[-2\left(2n_r+\ell+\frac{D}{2}\right)+\sqrt{1+\frac{8\mu V_0}{\hbar^2\alpha^2}}\right]  \eqno{(50)}$$ 
Thus by HFT, we have
$$\langle r^{-2}\rangle=\frac{\alpha^2}{(2\ell+D-2)}\left[-2\left(2n_r+\ell+\frac{D}{2}\right)+\sqrt{1+\frac{8\mu V_0}{\hbar^2\alpha^2}}\right]  \eqno{(51)}$$
Similarly, putting $q=V_0$ in Eq.(47), we obtain $\langle V(r)\rangle$ as $$\langle V(r)\rangle=V_0\left(1+\frac{8\mu V_0}{\hbar^2\alpha^2}\right)^{-\frac{1}{2}}\left[-2\left(2n_r+\ell+\frac{D}{2}\right)+\sqrt{1+\frac{8\mu V_0}{\hbar^2\alpha^2}}\right]  \eqno{(52)} $$
Finally, we note that the expectation value $\langle T\rangle$ can also be obtain from the fact that $E_{n_r}=\langle T\rangle+\langle V\rangle$.

\pagebreak
\noindent{\bf 6. Conclusions.}\\\\
In this paper, we present the solutions to the  $D$-dimensional Schr$\ddot{o}$dinger equation with the modified P$\ddot{o}$schl-Teller potential within the framework of an approximation to the centrifugal term. The energy eigenvalues were obtained in $D$-dimensions using the Nikiforov-Uvarov method, and it was found to agree with previous works [2, 24]. The corresponding eigenfunctions of the modified P$\ddot{o}$schl-Teller Potential were obtained in terms of the Jacobi polynomials. The energy levels for some neighboring dimensions were found to be getting closer as for higher exited states. Moreover, the graph of the absolute value of the unnormalized hyperradial function was plotted against  $r$ for some $\ell\neq 0$ states in 3$D$ and 5$D$ (Figure 2-5), and the curves become steeper with increase in the parameter $\alpha$.  
 
Finally, the normalization constants were obtained in form of the hypergeometric series and some useful expectation values were computed using the Feynman-Hellmann theorem.

\pagebreak

\pagebreak

\centering {\bf FIGURES}.

In the figure, $U_{n\ell}(r)=r^{-(D-1)/2}R_{n_r,\ell}(r)$,$V_0=\mu=\hbar=1.$
\begin{figure}[ht]
\centering

\includegraphics[scale=.52]{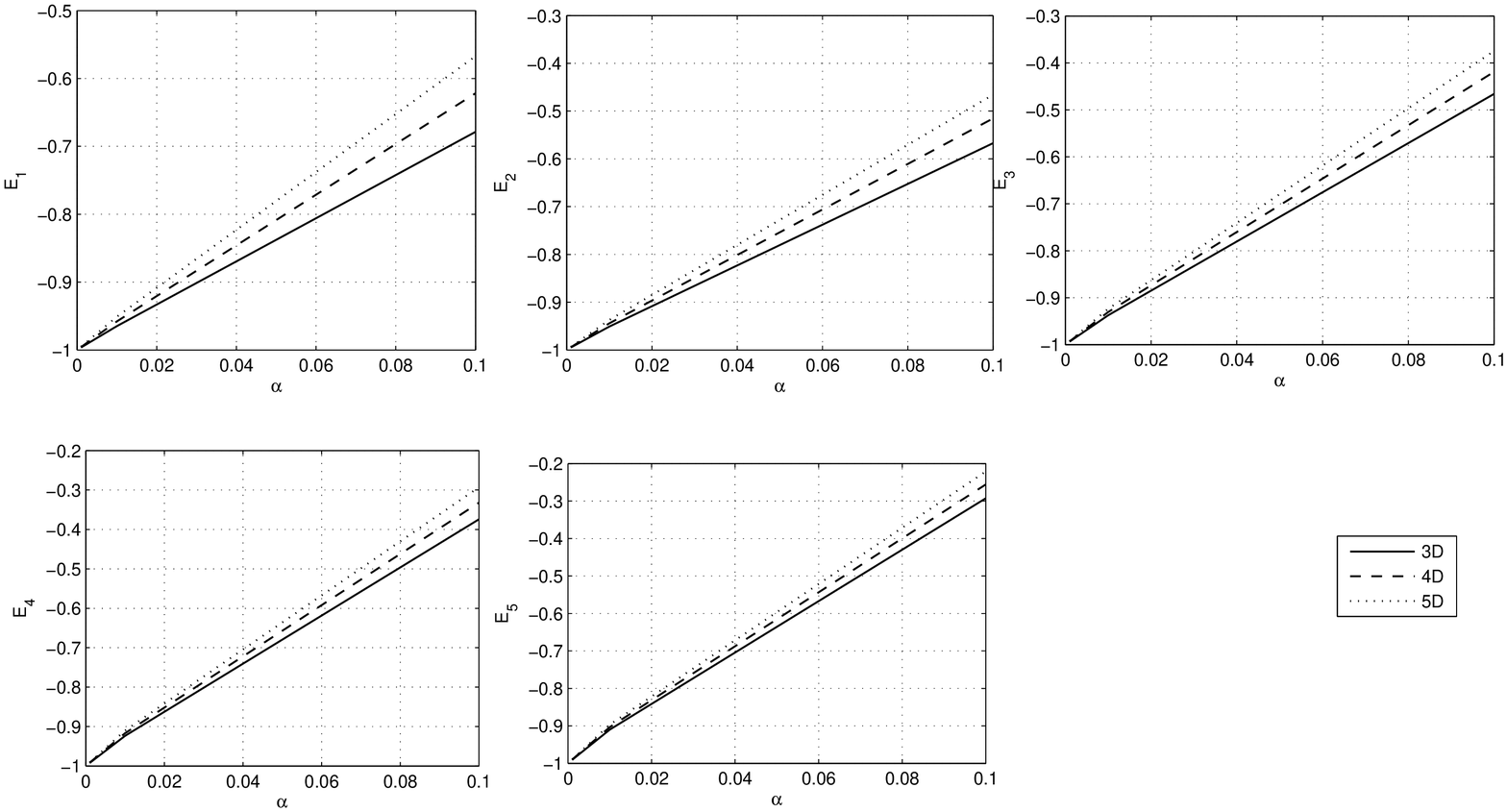}
\caption{$E_n$ against $\alpha$ for some dimensions.}
\label{fig:}
\end{figure}
\hfil

\begin{figure}[t]
\centering

\includegraphics[scale=.6]{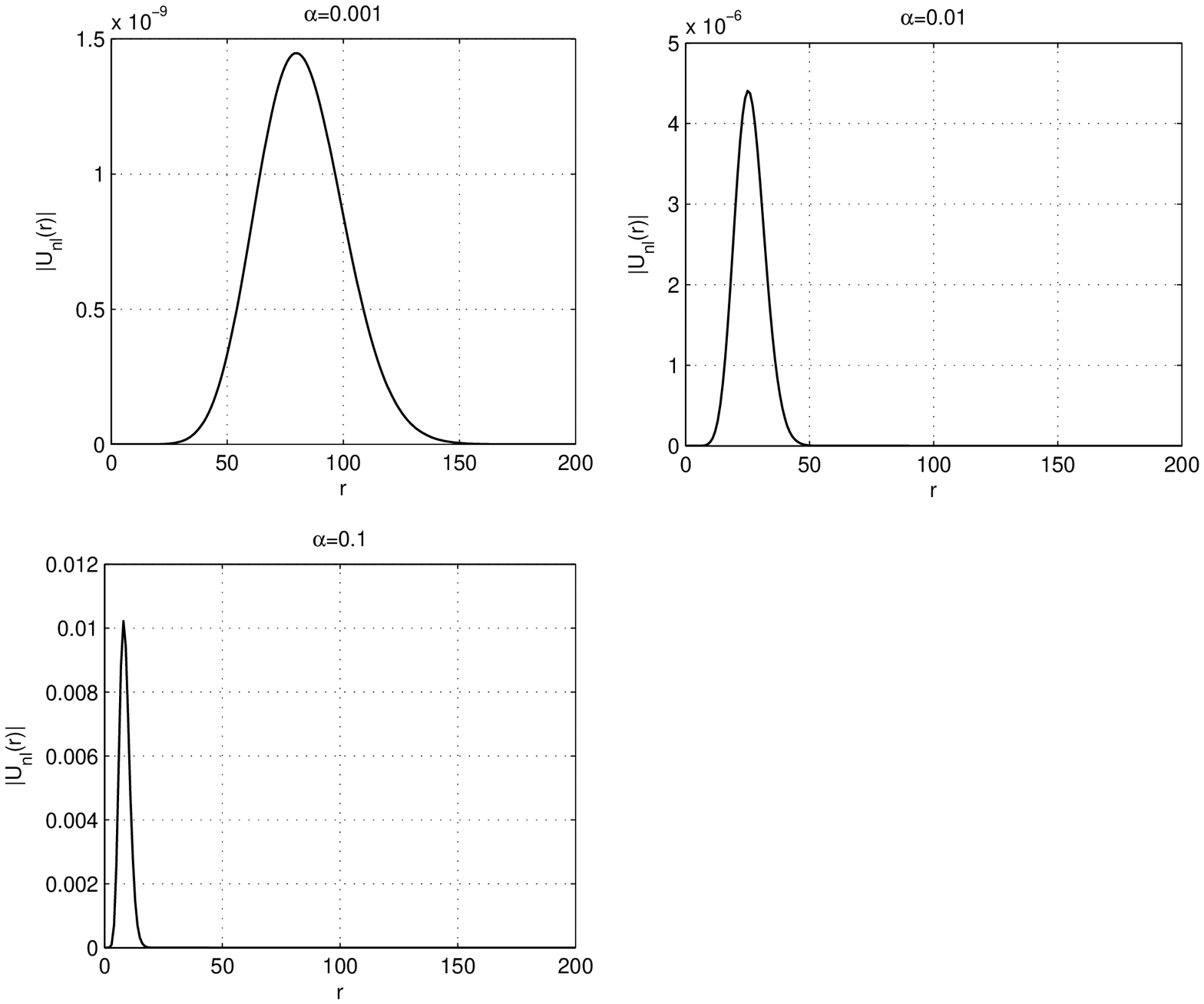}
\caption{$|U_{{n_r}\ell}(r)|$ against $r$ for $n_r=0,\ell=1$ in 3-dimensions.}
\label{fig:}
\end{figure}
\hfil
\begin{figure}[t]
\centering

\includegraphics[scale=.6]{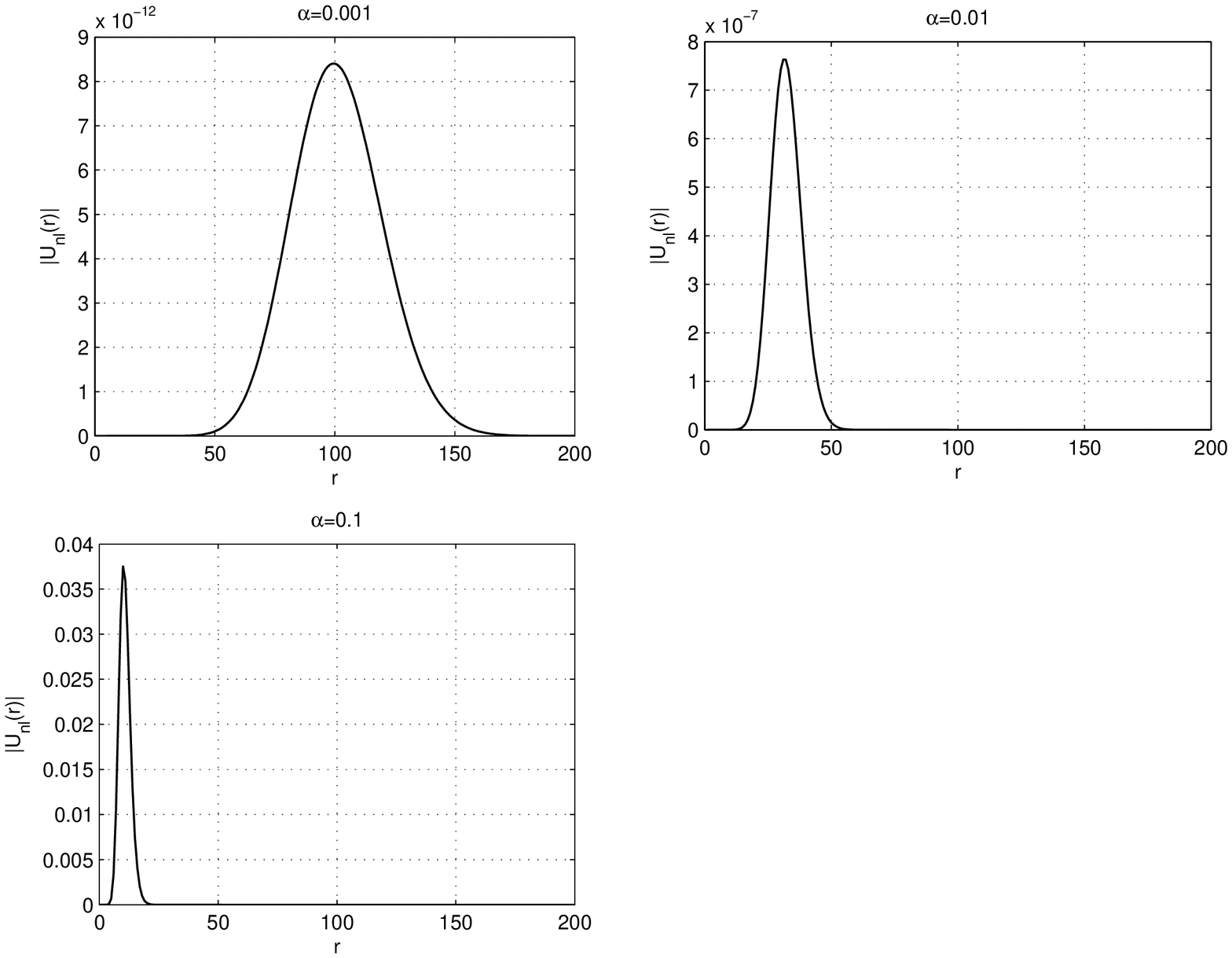}
\caption{$|U_{{n_r}\ell}(r)|$ against $r$ for $n_r=0,\ell=1$ in 5-dimensions.}
\label{fig:}
\end{figure}
\hfil
\begin{figure}[t]
\centering
\includegraphics[scale=.6]{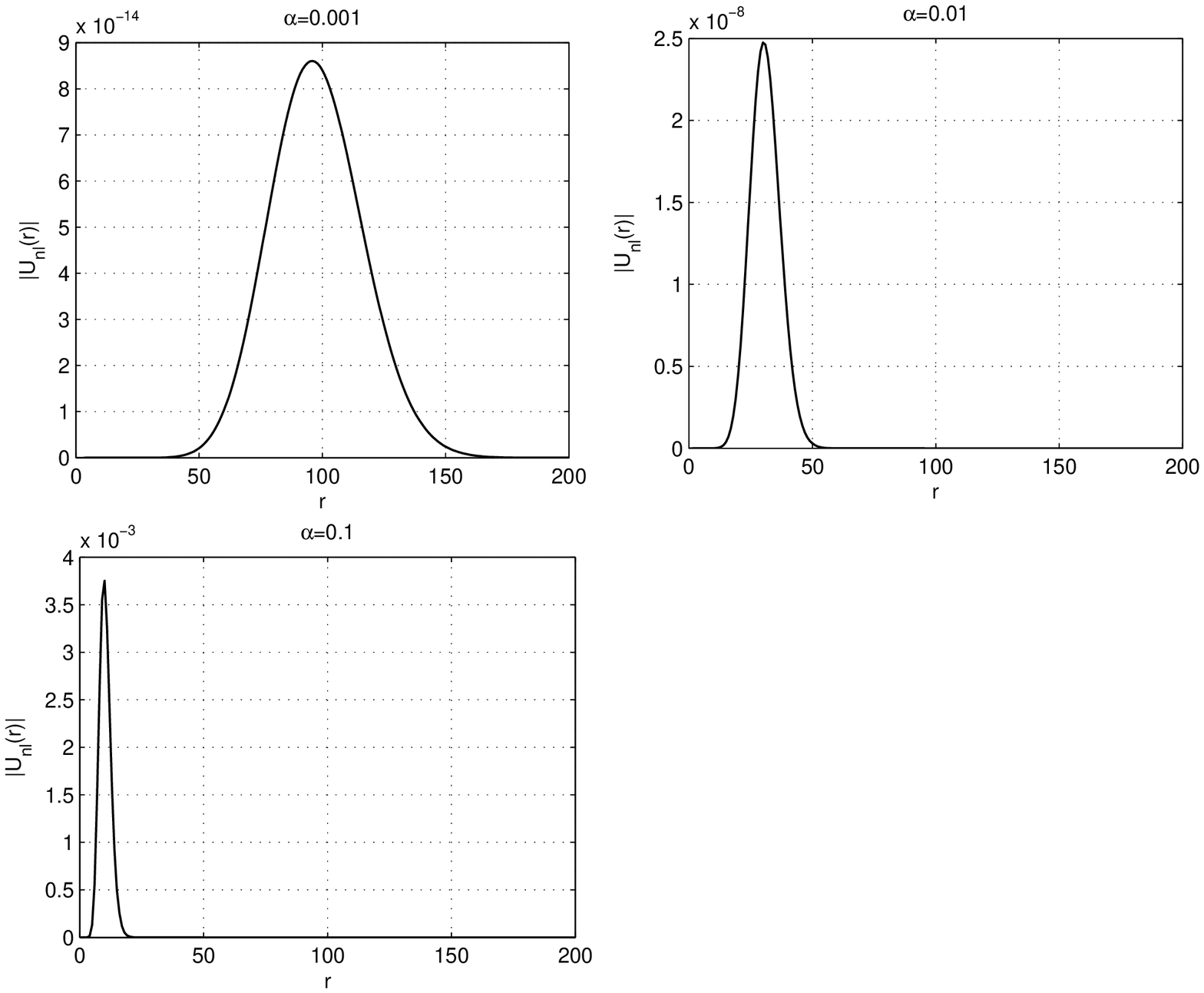}
\caption{$|U_{{n_r}\ell}(r)|$ against $r$ for $n_r=0,\ell=2$ in 3-dimensions.}
\label{fig:}
\end{figure}
\hfil
\begin{figure}[t]
\centering
\includegraphics[scale=.6]{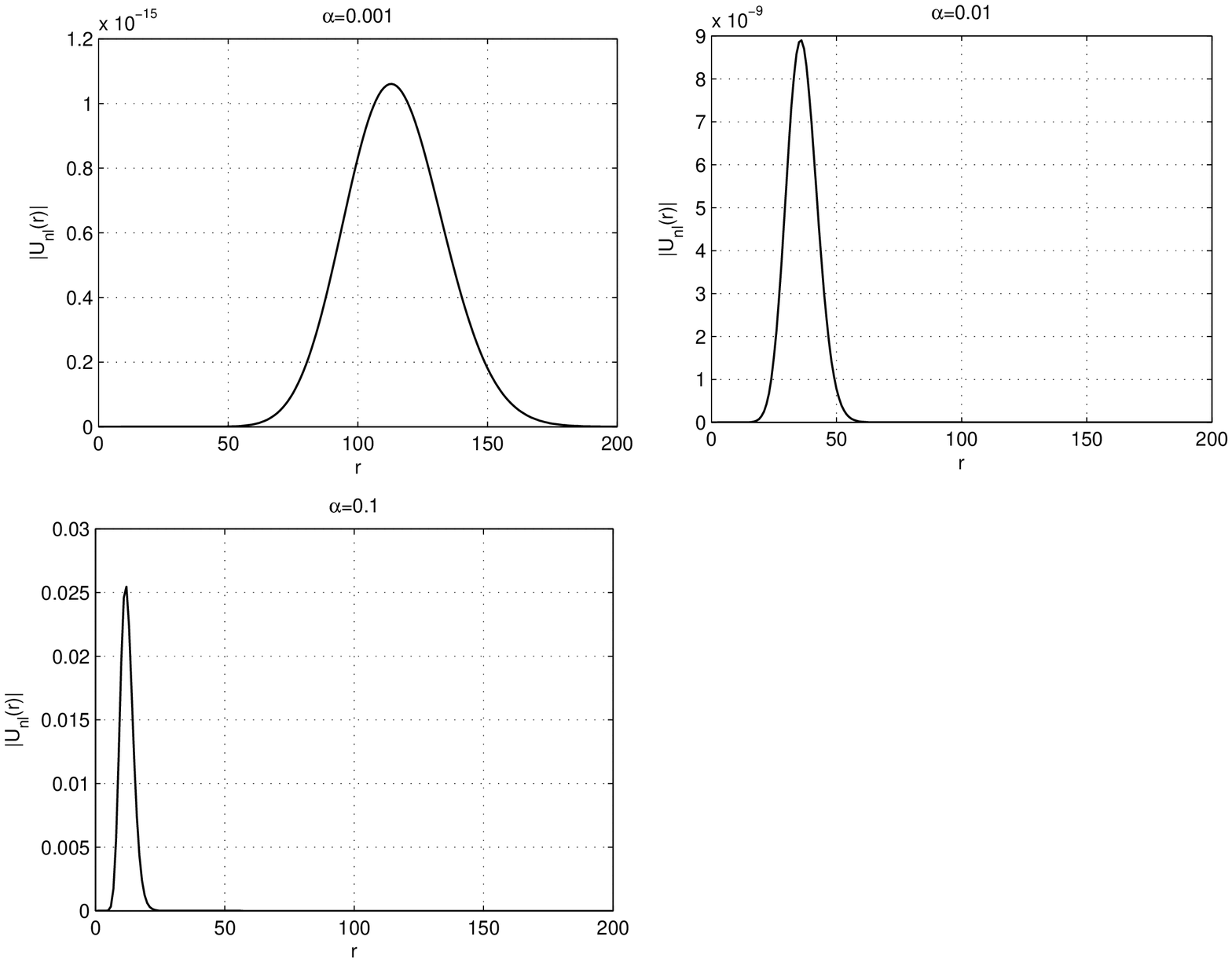}
\caption{$|U_{{n_r}\ell}(r)|$ against $r$ for $n_r=0,\ell=2$ in 5-dimensions.}
\label{fig:}
\end{figure}
\end{document}